%% file: main.tex
\documentclass[times,twocolumn,final]{elsarticle}

\usepackage{cag}
\usepackage{framed,multirow}

\usepackage{amssymb}
\usepackage{amsmath}
\usepackage{latexsym}

\usepackage{url}
\usepackage{xcolor} 
\usepackage{hyperref}

\usepackage[caption=false,font=footnotesize]{subfig}
\usepackage[switch,pagewise]{lineno}
\usepackage{algorithm}
\usepackage{algpseudocode}
\usepackage{booktabs}
\usepackage{float}
\usepackage{enumitem}

\definecolor{myred}{rgb}{1,0.3,0.3}

\journal{Computers \& Graphics}

\begin{document}

\verso{Preprint Submitted for review}

\begin{frontmatter}

\title{Interactive Exploration of Large-scale Streamlines of Vector Fields via a Curve Segment Neighborhood Graph}

\author[1]{Nguyen K.\ Phan\corref{cor1}}
\author[2]{Brian Kim}
\author[1]{Adeel Zafar}
\author[1]{Guoning Chen}
\cortext[cor1]{Corresponding author: nguyenpkk95@gmail.com}
\address[1]{University of Houston}
\address[2]{Klein Cain High School}

\received{\today}

\begin{abstract}
\input{Content/abstract}

\end{abstract}


\end{frontmatter}


\input{Content/introduction}
\input{Content/related_work}
\input{Content/our_method}
\input{Content/results}

\input{Content/conclusion}


\bibliographystyle{cag-num-names}
\bibliography{Content/references}

\end{document}

%% file: Content/abstract.tex
Streamlines have been widely used to represent and analyze various steady vector fields.
To sufficiently represent important features in complex vector fields (like flow), a large number of streamlines are required.
Due to the lack of a rigorous definition of features or patterns in streamlines, user interaction and exploration are required to achieve effective interpretation.
Existing approaches based on clustering or pattern search, while valuable for specific analysis tasks, often face challenges in supporting interactive and level-of-detail exploration of large-scale curve-based data, particularly when real-time parameter adjustment and iterative refinement are needed.
To address this, we design and implement an interactive web-based system. Our system utilizes a Curve Segment Neighborhood Graph (CSNG) to encode the neighboring relationships between curve segments. CSNG enables us to adapt a fast community detection algorithm to identify coherent flow structures and spatial groupings in the streamlines interactively. CSNG also supports a multi-level exploration through an enhanced force-directed layout. Furthermore, our system integrates an adjacency matrix representation to reveal detailed inter-relations among segments. To achieve real-time performance within a web browser, our system employs matrix compression for memory-efficient CSNG storage and parallel processing. We have applied our system to analyze and interpret complex patterns in several streamline datasets. Our experiments show that we achieve real-time performance on datasets with hundreds of thousands of segments.


%% file: Content/Introduction.tex
\section{Introduction}
\label{sec:intro}

Analyzing streamlines from velocity and vorticity fields is crucial for understanding complex steady-state flow behaviors in aerospace engineering, climate study, and energy harvesting. However, the sheer number of curves and their intricate relationships often lead to visual clutter, making it challenging to identify meaningful patterns and features.
To address this, clustering techniques \cite{YuHierarchicalClustering,BloodFlow2014,nguyen2021physics,shi2021integral} and pattern search methods \cite{lu2013exploring,FlowString2016,Wang2016Patterns} group similar curves or segments. While effective in many applications, they face challenges in large-scale interactive exploration. Clustering methods require expensive pairwise distance calculations, limiting real-time use. Pattern search approaches need manual template specification, hindering exploratory analysis. High computational costs (often minutes to hours for large datasets) make iterative parameter tuning difficult, preventing rapid experimentation with different settings.


To address these challenges, a system satisfying the following requirements is desired.

\begin{itemize}
\itemsep0em 
\item \textbf{R1:} It should support multi-level exploration of input streamlines, as features/patterns exist at different scales.
\item \textbf{R2:} It should achieve real-time performance (e.g., interactive frame rates $\geq$ 30 FPS for visualization with reasonable computation time for pattern extraction operations, including clustering and community detection) for large-scale datasets in web browsers. Features/patterns in streamlines are not always well-defined, requiring constant parameter tuning and exploration. This demands efficient pattern extraction and rendering despite limited memory, processing power, and the lack of GPU compute shaders in web browser environments.
\item \textbf{R3:} It needs intuitive user interaction to support parameter tuning and features/patterns exploration.
\end{itemize}

Recently, Phan and Chen \cite{phan2024curve} introduced the Curve Segment Neighborhood Graph (CSNG), a novel representation of streamlines that enables the application of community detection algorithms to group similar segments. This approach shows promise in extracting patterns from complex flow data. However, it lacks real-time performance for even small datasets and cannot handle large-scale data. Additionally, its CSNG visualization does not reveal detailed spatial configurations of individual segments.


In this paper, we present a new interactive system that addresses the above requirements for the effective exploration of large-scale curve-based vector field data (specifically streamlines). Our system improves the work \cite{phan2024curve} with the following unique features.


\begin{itemize}[leftmargin=10pt]
\itemsep0em 

\item Our system supports CSNG construction describing relations between streamlines (coarsest level), between sub-curves (of streamlines) decomposed by curvature \cite{Li:2015:EFF:2838431.2838604} (middle level), and between individual segments (finest level). Community detection applies to any level (addressing \textbf{R1}). Classical clustering (e.g., PCA $k$-means) is also supported.

\item Our web-based system supports hierarchical community exploration via enhanced force-directed layout, enabling automated feature detection and manual refinement through interactive splitting and merging of graph communities (addressing \textbf{R1} and \textbf{R3}). To aid the inspection of the granular segment-to-segment configurations within individual communities, an adjacency matrix, called the Adjacency Matrix of Curve Segments (AMCS), for the sub-graph of a community, is visualized (addressing \textbf{R3}). 


\item Our system improves performance through efficient parallel processing and memory management. Utilizing SharedArrayBuffer for zero-copy data sharing between web workers, we eliminate memory duplication during multithreaded operations like neighbor search, community detection, and matrix visualization (addressing \textbf{R2}).



\end{itemize}


We show that community detection using the Louvain algorithm \cite{blondel2008fast} on CSNGs achieves real-time performance on large-scale curve datasets, operating orders of magnitude faster than traditional clustering with comparable or improved quality. Quality is measured through visual coherence analysis and, where available, quantitative metrics such as the Jaccard index against ground truth labels (achieving 0.712 with our method vs. 0.275 with PCA k-means on the Couette flow dataset; see Section~\ref{subsec:comparative}).

We demonstrate effectiveness through case studies on streamline datasets, showcasing its ability to extract meaningful patterns from complex flow data. We demonstrate scalability through the successful analysis of the turbulent Couette flow dataset \cite{li2019direct,zafar2023extract} with 1.4 million curve segments, showing effective handling of extreme density variations in vortex bundles while maintaining interactive performance.




%% file: Content/related_work.tex
\section{Related Work}
\label{sec:relatedWork}

\subsection{Streamline placement and visualization}

Streamlines are commonly used to visualize flow data \cite{GeomSTAR09,STARSurfVis12}. Effective streamline representation requires seed selection and placement (seeding strategy). Numerous strategies \cite{sane2020survey} achieve evenly-spaced streamlines in physical \cite{jobard1997creating,liu2006advanced,mebarki2005farthest,chen2007similarity} or image space \cite{turk1996image,spencer2009evenly,li2007image}, produce feature-enhanced placements \cite{verma2000flow,chen07,wu2009topology}, or seed streamlines based on information theory \cite{FlowInfo10}. Parallel and distributed computing strategies address large-scale flow data \cite{zhang2018survey}.





To visualize computed streamlines and aid flow interpretation, rendering, processing, and user interaction techniques have been proposed.
G\"unther et al. \cite{gunther2013opacity} optimized opacity values for individual curve segments to reveal inner patterns occluded by dense 3D streamline placement.
Tong et al. \cite{tong2015view} introduced view-dependent streamline deformation to unveil hidden structures.
Lu et al. \cite{lu2021curve} introduced a KD-tree data structure for curve segments to support interactive exploration and querying of 3D curves. Viewpoint selection techniques \cite{LeeMSC11,tao2012unified} automatically select viewpoints for streamline rendering to better depict flow patterns.

\subsection{Streamline exploration}

Feature extraction methods in flow visualization can be generally categorized into detecting mathematically well-defined features, such as vector field topology and vortex cores \cite{post2002feature,post2003state}, and identifying features that are not always mathematically well-defined, such as vortex regions and other coherent structures. Our work focuses on the latter, especially patterns in sets of streamlines. Existing approaches for analyzing these ill-defined features or patterns largely fall into two groups: clustering and pattern search. Clustering methods partition the data into groups, whereas pattern search methods filter and select curves or sub-curves of interest. Our approach belongs to the clustering category.


Techniques for clustering input curves have been proposed to aid blood flow analysis \cite{BloodFlow2014,meuschke2016clustering,meuschke2018exploration} and fiber track analysis in diffusion tensor imaging \cite{moberts2005evaluation,everts2015exploration}. They apply not only to streamlines \cite{shi2021integral} but also to pathlines \cite{nguyen2021physics}. Both classic clustering methods and machine learning-based clustering \cite{han2018flownet} have been introduced. 
Many clustering methods require selecting specific similarity measures and combining them with clustering algorithms (see the survey by Shi et al. \cite{shi2021integral}). These methods are often computationally intensive and do not fully support the requirements of multi-level exploration (\textbf{R1}) and real-time performance (\textbf{R2}). For example, Agglomerative Hierarchical Clustering (AHC) \cite{YuHierarchicalClustering,nguyen2021physics} supports multi-level exploration (\textbf{R1}) but does not meet the real-time performance requirement (\textbf{R2}). Other clustering techniques typically do not support \textbf{R1} and likely fail to meet \textbf{R2}. Our method addresses these limitations.


Streamline embedding maps streamlines into a low-dimensional space as an alternative to direct clustering.
R\"ossl and Theisel \cite{rossl2011streamline} embedded streamlines as points in feature space where spatial proximity approximates Hausdorff distance, yielding a manifold representation for classifying global stream behaviors. While they focused on continuous manifold representation of whole streamlines, our work uses a discrete graph-based representation (CSNG) modeling topological neighborhoods at the segment level, enabling local flow community detection through modularity optimization.

Pattern search identifies curves or segments similar to user-specified references. Wang et al. \cite{wang2014pattern, wang2015multi, Wang2016Patterns} developed pattern identification methodologies for vector field data. Lu et al. \cite{lu2013exploring} introduced distribution-based streamline characterization. Tao et al. \cite{Flowstring2014,FlowString2016} encoded streamline characteristics into character strings to facilitate pattern search. Pattern search strategy requires user-specified references as input and may miss important patterns that the user has no knowledge of.

Graph-based approaches have been introduced to aid 3D flow field exploration and analysis. Xu et al. introduced Flow Web \cite{xu2010flow}, representing field regions as nodes connected by particle travel. Xu and Shen proposed FlowGraph \cite{ma2013flowgraph}, organizing streamline clusters and spatial regions hierarchically. Phan and Chen \cite{phan2024curve} introduced the Curve Segment Neighborhood Graph (CSNG) to represent neighboring relations among segments, enabling graph segmentation techniques like community detection for pattern extraction and exploration.
Our work adapts the idea of graph representation to assist the grouping (or clustering) of segments and curves. Compared to \cite{phan2024curve}, our work extends CSNG to represent relations among individual curves, sub-curves, and segments. We address performance issues using matrix compression and zero-copy data sharing, and incorporate additional exploration functionality for multi-level data exploration.


\subsection{Community Detection and Large Graph Visualization}

Community detection methods have been extensively applied in various fields, but not yet in curve-based data analysis. Fortunato et al. \cite{fortunato2010community} and Lancichinetti et al. \cite{lancichinetti2009community} reviewed community detection methods. Large graph visualization techniques include force-directed layouts \cite{liu2022improved, govyadinov2019graph} and navigation strategies \cite{eades2004navigating}.


Adjacency matrix visualization is challenging for large, sparse matrices. Dinkla et al. \cite{dinkla2012compressed} combined node-link diagrams with adjacency matrices and introduced compression. Henry et al. \cite{henry2007matlink} enhanced interactivity by highlighting connections on matrix headers. Our work builds upon these foundations to address complexities in navigating our expansive adjacency matrix.

%% file: Content/our_method.tex
\section{Design of Our System}


Our system is designed for exploratory analysis of curve-based flow data, discovering patterns that may not be mathematically well-defined. This complements automatic feature extraction methods, which excel at identifying well-defined features (e.g., critical points, vortex cores) but may struggle with application-specific patterns that require domain expertise. Real-time performance enables iterative exploration, where users can rapidly test hypotheses, validate automatically detected patterns, and refine results based on domain knowledge. Our method involves design choices and configurable parameters (distance measures, neighborhood strategies, segmentation approaches, and community detection settings). This provides flexibility to adapt to different analysis scenarios, requiring users to understand when to apply each option. We provide default recommendations and guidance on parameter selection throughout this section, recognizing that optimal settings may vary depending on the specific dataset and analysis goals.

\begin{figure}[!t]
   \centering
   \includegraphics[width=0.98\linewidth]{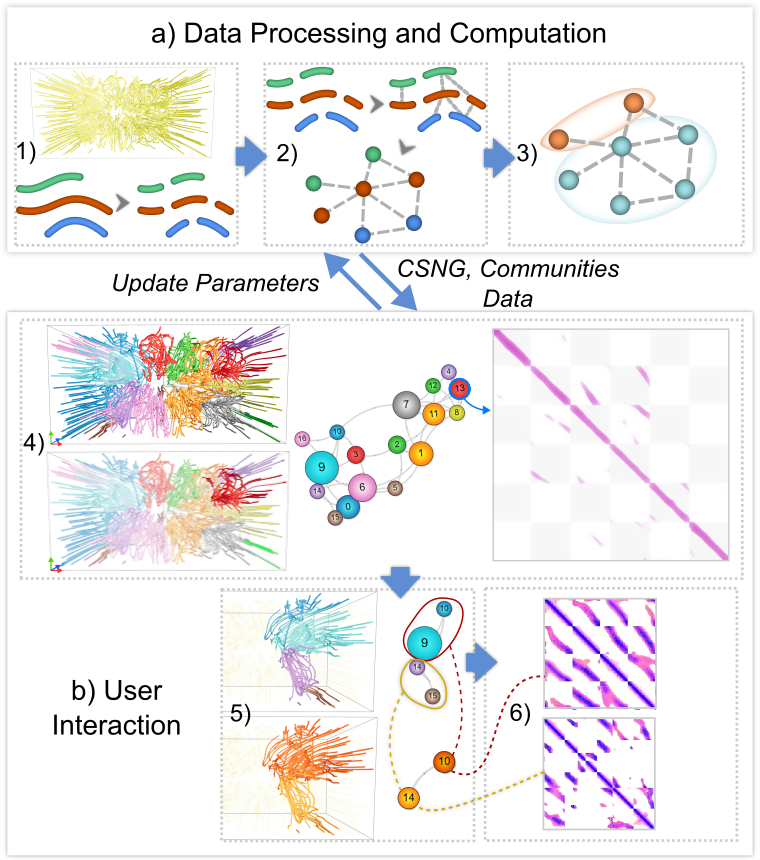}
   \caption{Illustration of our framework. a) Data processing and computation component consists of 1) streamline decomposition, 2)  CSNG construction, and 3) community detection on the obtained CSNG. b) User interaction supports 4) multi-view exploration and 5) manual adjustments of the community results, such as merging and splitting communities to fix misclassification using the graph view.
   }
   \label{fig:pipeline}
\end{figure}

Figure~\ref{fig:pipeline} illustrates our system design with two main components: a) data processing and computation, and b) user interaction. Data processing constructs CSNG from input streamlines and extracts communities. User interaction supports multi-level exploration through multiple visual representations, including parameter tuning for detecting new communities.

The data processing component addresses \textbf{R1} by constructing multi-level CSNGs (segment, sub-curve, and streamline levels) and addresses \textbf{R2} through fast community detection algorithms and parallel processing optimizations. The user interaction component addresses \textbf{R1} and \textbf{R3} via synchronized multi-view visualizations (3D spatial view, force-directed graph view, and adjacency matrix view) with interactive refinement, enabling both automated pattern detection and manual adjustment based on domain knowledge. These components form a feedback loop where user interactions trigger CSNG and community recomputation, with results immediately reflected in all views.


In the subsequent sections, we provide more details on these steps and functionalities, and their relations. 

\section{Data Processing and Computation}

This section details the construction of CSNG and community detection from CSNG (Figure~\ref{fig:pipeline}a).

\subsection{Curve Decomposition and CSNG Construction}
\label{sec:multilevel_explore}

To address \textbf{R1}, we adopt the \emph{Curve Segment Neighborhood Graph (CSNG)} introduced by Phan and Chen \cite{phan2024curve}. The input consists of streamlines represented as polylines (sequences of 3D points), typically obtained from computational fluid dynamics CFD solvers. During preprocessing, streamlines can optionally be resampled for uniform point spacing, and very short streamlines may be filtered based on user-specified thresholds.

A CSNG is defined as a directed graph $G_{CS} = (V_{CS}, E_{CS})$, where:
\begin{itemize}
  \setlength{\itemsep}{0pt}
  \setlength{\parskip}{0pt}
\item $V_{CS}$ represents the set of nodes, each corresponding to a curve segment
\item $E_{CS}$ denotes the set of directed edges indicating neighbor relationships
\item Node attributes can include segment properties such as curvature, length, and velocity magnitude
\item Edge attributes can store property differences between neighboring segments
\end{itemize}
\emph{Segments} are individual line segments (edges connecting consecutive points) on streamlines, while \emph{sub-curves} are groups of consecutive segments. We employ a simplified approach where each sub-curve merges $N$ consecutive segments. While more robust decomposition methods using geometric attributes (e.g., curvature variation, torsion changes) exist, our fixed-length segmentation offers computational efficiency and predictable behavior, well-suited for our interactive exploration framework.

\paragraph{Multi-level CSNG}
In this work, we extend CSNG to represent the neighboring relations among the individual streamlines and the individual sub-curves, respectively, to achieve a multi-level representation of the neighborhood relations. 
In the neighborhood graph for streamlines, each node represents a streamline, and an edge connecting two nodes indicates that the two corresponding streamlines are neighboring each other. 
In the neighborhood graph for the sub-curves that are decomposed using the curvature information \cite{Li:2015:EFF:2838431.2838604}, each node corresponds to a sub-curve, and an edge connecting two nodes means that the two corresponding sub-curves are neighboring each other. 

At both levels of the neighborhood graph, we can store various attributes as node and edge weights to enrich the analysis. These attributes serve multiple purposes: (1) weights in community detection algorithms to favor grouping geometrically similar elements, (2) filtering and selection of communities based on flow characteristics, and (3) additional visual encoding options for exploration.

For nodes, we store properties, including:
\begin{itemize}
  \setlength{\itemsep}{0pt}
  \setlength{\parskip}{0pt}
\item Saliency: average orientation difference with neighboring segments, quantifying how much a segment stands out~\cite{chen2007similarity}. Higher saliency values indicate regions where flow direction changes significantly, often corresponding to boundary layers, separation regions, or transition zones.
\item Average distance to neighbors: reveals flow characteristics such as expansion or compression. Large distances may indicate flow separation or divergence, while small distances suggest converging or parallel flow structures.
\item Local curvature: rate of tangent direction change, useful for identifying vortical structures and recirculation zones. For highly curved or circular patterns, we compute curvature using a sliding window approach over a local neighborhood of points.
\end{itemize}

For edges, we store pairwise information, such as:
\begin{itemize}
  \setlength{\itemsep}{0pt}
  \setlength{\parskip}{0pt}
\item Direct distance between elements
\item Differences in curvature between connected elements
\end{itemize}

\paragraph{Determine Neighborhood Relationship}
To determine the neighborhood relations between streamlines, sub-curves, and individual segments, respectively, we consider two \textbf{neighbor search strategies}, i.e., $k$-nearest neighbor (kNN) and radius-based neighbor (RBN) \cite{phan2024curve}. The kNN approach identifies the top $k$ segments with the smallest distances to the query segment, while RBN selects all segments within a specified radius $R$. To optimize search efficiency, particularly for large datasets, we employ a segment-based KD-tree \cite{arya1998optimal} data structure for organizing segments and accelerating candidate identification.

The \textbf{distance measures} (or proximity functions) for determining neighborhood relations between segments include shortest distance, longest distance (or Hausdorff distance), and average distance. These are not formal distance metrics (may not satisfy the triangle inequality), but are practical proximity measures. By default, the longest distance is used, as it tends to select neighbors closer to the query curve \cite{phan2024curve}. Distance measure choice depends on analysis goals: shortest distance for detecting nearly parallel curves; longest distance for overall closeness and robustness to outliers; average distance for a balanced measure. We use consistent distance measures across all levels (segments, sub-curves, and streamlines) by default.

For streamlines and sub-curves, we determine their neighborhood relations using similar principles but adapted to their scales. Sub-curves are decomposed into streamlines where we group segments together, either using simple grouping approaches or more sophisticated curvature-based decomposition methods. Specifically:

\begin{itemize}
  \setlength{\itemsep}{0pt}
  \setlength{\parskip}{0pt}
\item For streamlines, distance is the minimum distance between any pair of sampled points. Streamlines are neighbors if this distance falls within top-$k$ distances (kNN) or within radius R (RBN).
\item For sub-curves decomposed based on curvature information, we compute the distance between two sub-curves using the Hausdorff distance (longest distance) between the curve segments, similar to the approach used for streamlines. This distance metric considers all point pairs between the two sub-curves to find the maximum minimum distance. Neighborhood relations are then established using either kNN or RBN approaches with these distance metrics.
\end{itemize}

This hierarchical approach to determining neighborhoods maintains consistency in our distance metrics across levels.

\subsection{Pattern Extraction via Community Detection}
\label{sec:communitydetect}


The next step in our analysis pipeline extracts meaningful patterns from curve segments in our CSNG. Traditional clustering techniques like DBSCAN, $k$-means, and agglomerative hierarchical clustering have been widely used for curve analysis \cite{shi2021integral}. However, these methods rely on pairwise distance calculations, becoming computationally expensive for large datasets. They struggle to incorporate the rich relational information in our graph structure and may require extensive parameter tuning.
Traditional clustering methods operate on feature vectors or distance matrices, processing each segment independently based on geometric similarity (e.g., distance in PCA-reduced feature space for $k$-means, density reachability for DBSCAN). In contrast, CSNG explicitly encodes spatial neighborhood topology as a graph structure with edges representing adjacency relationships. Graph-based community detection algorithms like Louvain leverage this topological information by optimizing modularity, considering not just pairwise similarities but the network structure of connections. This enables identifying cohesive groups with dense internal connections but sparse external connections, a pattern that traditional clustering cannot directly exploit. Regarding parameter sensitivity, distance-based clustering requires careful selection of multiple parameters (e.g., $k_c$ for $k$-means, $\epsilon$ and minPts for DBSCAN, linkage criteria and cut height for hierarchical clustering), significantly affecting results \cite{YuHierarchicalClustering,shi2021integral}. While community detection also requires parameters (e.g., resolution in Louvain), it operates on the already-constructed neighborhood graph, making it more robust to variations in underlying distance metrics.

To address these limitations, we implement two variations of community detection using the Louvain algorithm \cite{blondel2008fast}: segment-based detection on the original CSNG and streamline-based detection on aggregated data, accommodating multi-level CSNGs. Both optimize modularity while offering different granularity in pattern identification.

\paragraph{Segment-based Community Detection}
In the segment-based approach, we apply the Louvain algorithm directly to the CSNG, where each node represents an individual curve segment. Modularity optimization is performed according to \cite{blondel2008fast}:

\begin{equation}
    M_s = \frac{1}{2W} \sum_{i=1}^{N_s}{\sum_{j=1}^{N_s}{\left(\left[w_{ij} - \frac{W_iW_j}{2W}\right]\delta(c_i, c_j)\right)}}
    \label{eq:seg_modularity}
\end{equation}

\noindent where $N_s$ is the total number of segments, $w_{ij}$ is the edge weight between segments $i$ and $j$ ($w_{ij} = 1$ if an edge exists, otherwise $w_{ij} = 0$), $W_i$ and $W_j$ are sums of edge weights adjacent to segments $i$ and $j$, $W$ is the total sum of all edge weights, $c_i$ and $c_j$ represent community assignments (labels) of segments $i$ and $j$ (e.g., $c_i = 3$ means segment $i$ belongs to community 3), and $\delta$ is the Kronecker delta function ($\delta(c_i, c_j) = 1$ if $c_i = c_j$, otherwise 0, ensuring we only count edges within the same community).

This modularity function measures community partition quality by comparing the edge density within communities to the expected edge density in a random graph. Higher modularity scores (ranging from -1 to 1) indicate stronger community structure, meaning segments within the same community are more densely connected than expected by chance.

\paragraph{Streamline-based Community Detection}
For the neighborhood graph built from individual streamlines, we apply the Louvain algorithm directly, achieving significantly faster community detection with lower memory requirements, as we only process and store relationships between streamlines rather than individual segments. While offering computational advantages, this sacrifices the finer granularity of local segment information that segment-based detection provides. This makes streamline-based community detection suitable for initial exploration of large datasets, with the option to switch to segment-based detection for detailed analysis of regions of interest.

The streamline-based approach first aggregates segment-level relationships to create a streamline neighborhood graph. For each streamline pair $(S_a, S_b)$, we compute relationship strength based on constituent segments:

\begin{equation}
    R(S_a, S_b) = \frac{\sum_{i \in S_a}{\sum_{j \in S_b}{w_{ij}}}}{|S_a| \cdot |S_b|}
    \label{eq:streamline_relation}
\end{equation}

\noindent where $|S_a|$ and $|S_b|$ denote segment counts in streamlines $S_a$ and $S_b$. This normalized relationship strength accounts for varying streamline lengths.

The Louvain algorithm is then applied to this aggregated graph with the following modularity function:

\begin{equation}
    M_l = \frac{1}{2W_l} \sum_{i=1}^{N_l}{\sum_{j=1}^{N_l}{\left(\left[R_{ij} - \frac{W_i^lW_j^l}{2W_l}\right]\delta(c_i, c_j)\right)}}
    \label{eq:line_modularity}
\end{equation}

\noindent where $N_l$ is the total number of streamlines, $R_{ij}$ is the relationship strength between streamlines $i$ and $j$ (Equation (\ref{eq:streamline_relation})), $W_i^l$ and $W_j^l$ are sums of relationship strengths adjacent to streamlines $i$ and $j$, and $W_l$ is the total sum of all relationship strengths.

Note that community detection on the CSNG encoding neighborhood relations between \emph{sub-curves} can be achieved similarly to the above streamline-based community detection.

\begin{figure*}[!t]
    \centering
    \includegraphics[width=0.98\linewidth]{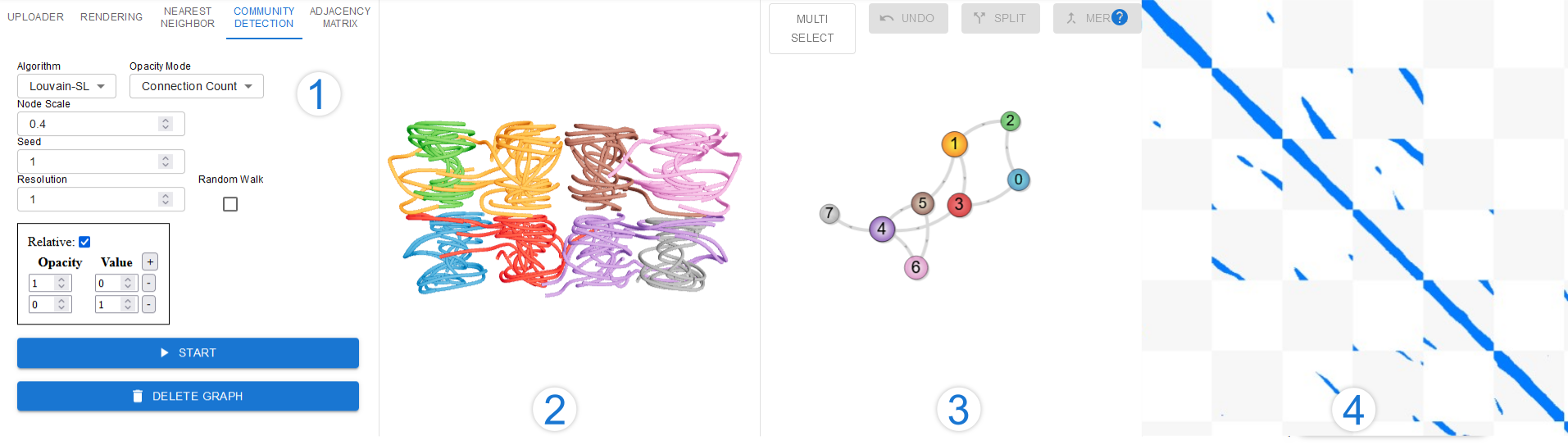}
    \caption{The main interface of our system consists of four key components: {\large\textcircled{\small 1}} Control panel for settings, currently shown algorithm parameters including community detection settings and visualization options, {\large\textcircled{\small 2}} 3D visualization view displaying the streamlines colored by their community assignments, {\large\textcircled{\small 3}} Force-directed graph visualization showing the community structure and relationships, {\large\textcircled{\small 4}} adjacency matrix view displaying detailed geometric patterns.}
    \label{fig:mainUI}
\end{figure*}

\paragraph{Comparative Advantages}
Segment-based detection provides finer granularity in pattern identification and better captures local flow features, as demonstrated in Figure~\ref{fig:crayfish_comparison}(d) where it successfully isolates a localized vortex feature. This approach allows detection of sub-streamline patterns and remains more sensitive to local geometric variations. In contrast, streamline-based detection offers more stable community assignments with reduced computational overhead (see Table~\ref{tab:combined_stats_performance} for timing comparisons), better preserves global flow structure as shown in Figure~\ref{fig:crayfish_comparison}(a), and demonstrates increased robustness to local noise and variations by averaging neighborhood relationships over entire streamlines, providing an overview of flow patterns at the streamline level.

The resolution parameter in both approaches controls detected community granularity, with smaller values leading to coarser partitioning and larger values resulting in finer partitioning. Users can select the appropriate approach based on analysis needs: segment-based detection favors local detail while streamline-based detection emphasizes global structure.

\section{User Interaction and Interface}
\label{sec:system}

This section describes the second major part of our pipeline: user interface and interaction mechanisms (illustrated in ~\autoref{fig:pipeline}b).

\begin{figure}[!t]
    \centering
    \includegraphics[width=0.98\linewidth]{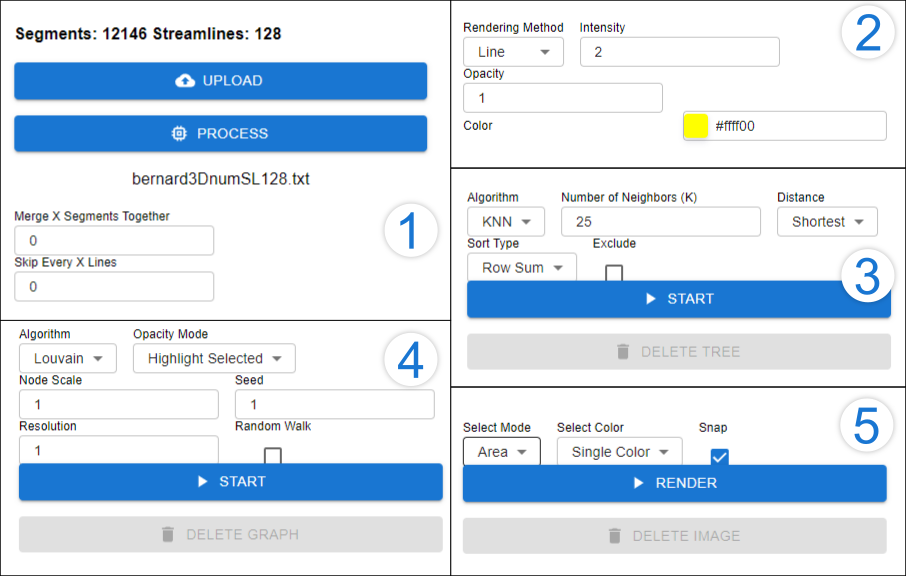}
    \caption{The five configuration panels in our system: {\large\textcircled{\small 1}} Dataset upload and preprocessing panel for loading streamline data and configuring merge/skip parameters, {\large\textcircled{\small 2}} Rendering settings panel for controlling visualization parameters including method, intensity, opacity and color, {\large\textcircled{\small 3}} Nearest neighbor configuration panel for setting algorithm type (kNN/RBN), number of neighbors, and distance metrics, {\large\textcircled{\small 4}} Community detection panel for configuring Louvain algorithm parameters like node scale and resolution, and {\large\textcircled{\small 5}} Adjacency matrix visualization panel for controlling display mode and color schemes.}
    \label{fig:settings}
\end{figure}

Our web-based system supports visualizations and user interactions. \autoref{fig:mainUI} shows our system with multiple linked-views. A 3D view displays streamlines colored by their corresponding communities. A graph view shows connectivity between communities for exploration and editing. Interactions and modifications in the graph view are reflected in the 3D view. A matrix view provides information on pairwise relations at streamline and segment levels.
Parameter setups for streamline decomposition, neighbor search for CSNG construction, Louvain community detection, and visualization effects are configured through panels shown in \autoref{fig:settings}.
In the following, we provide details on the design of these critical components (\autoref{subsec:userinterface}) and system implementation (\autoref{subsec:implementation}).


\subsection{User Interaction Components}
\label{subsec:userinterface}

\paragraph{Force-Directed Layout}
We use a multi-layered force-directed graph to visualize CSNG and its communities hierarchically.
Our force-directed graph is formulated as a compound graph \(G_f = (V_f, E_f)\), where \(V_f\) represents nodes, each corresponding to a distinct community of curve segments from community detection algorithms. The set \(E_f\) consists of edges connecting these nodes, with each edge \((v_i, v_j) \in E_f\) indicating a neighborhood relationship between segments in clusters \(v_i\) and \(v_j\). To encode hierarchical information, each node may contain sub-nodes corresponding to sub-communities, and each edge could be a collection of edges connecting corresponding sub-communities. Our edge spring force in multi-layered graphs is inspired by Lu et al.'s work \cite{liu2022improved}, and our node spring force in clustered graphs adapts the work by Eades et al. \cite{eades2004navigating}. A detailed algorithm is provided in the supplemental document. We visualize a node with a sphere whose size is determined by the number of segments within the corresponding community, receiving a distinct color based on its community ID. 
We use curved arcs rather than straight lines to depict edges between communities. This design choice reduces visual clutter by preventing edge overlaps and makes it easier to distinguish individual connections when multiple edges emanate from the same node. Arc curvature is automatically adjusted based on the number of edges between node pairs, improving graph layout readability.
A user can interactively split a node (i.e., a community) from \(V_f\) into smaller sub-nodes, transforming the node into a sub-graph representing more detailed subdivisions. The splitting process is initiated by selecting a node and applying the corresponding split parameters (e.g., community detection algorithm, min and max node size). Visually, the newly created sub-graph is distinguished by a dashed outline, clearly indicating its derivation from the parent node (\autoref{fig:demo}(c)(d)). This visual distinction helps users track changes and understand the graph's evolving structure. Additionally, we provide an option for collapsing these sub-graphs back into a single node, enhancing the graph's flexibility \cite{phan2024curve}.

\paragraph{Adjacency Matrix of Curve Segments (AMCS)}
While the force-directed graph effectively visualizes community structure and inter-community relationships, it may not reveal detailed patterns within individual communities. To address this, we introduce the Adjacency Matrix of Curve Segments (AMCS), a complementary visualization for detailed analysis of segment relationships.


An AMCS is constructed as an $N \times N$ square matrix, where $N$ represents the number of curve segments in the selected portion of the CSNG. Each entry $(i,j)$ indicates the neighboring relationship between segments $i$ and $j$, with a value of 1 indicating a neighbor relationship and 0 otherwise. By default, segments are organized so those from the same streamline are adjacent and sorted from beginning to end. The matrix is asymmetric for kNN-based CSNGs but symmetric for RBN-based ones, reflecting their distinct neighbor search strategies.

The AMCS reveals characteristic patterns corresponding to specific geometric configurations of streamlines. For example, diagonal line structures typically indicate parallel or nearly parallel streamlines running in the same direction, while off-diagonal blocks suggest spatial proximity between distinct streamlines. However, pattern interpretation requires caution: similar visual patterns can arise from different geometric configurations depending on segment ordering. For instance, both tightly wound helical structures and loosely parallel flows can produce diagonal patterns, distinguished primarily by the thickness and density of diagonal bands. The matrix ordering (grouping segments by streamline vs. other criteria) significantly affects which patterns become visible, making the AMCS a powerful but context-dependent analytical tool.

For typical dense curve datasets, visualizing the complete AMCS becomes impractical due to display space limitations and visual cluttering. The computational overhead for rendering and interaction, combined with the cognitive load of interpreting dense matrices, makes it challenging to work with full-scale representations. Therefore, we recommend visualizing AMCS for selected communities or clusters. This focused approach enables users to examine detailed segment relationships within communities of interest while maintaining interactive performance and reducing visual complexity.



\subsection{System Performance Optimizations}
\label{subsec:implementation}
To ensure responsive interaction with large-scale streamline datasets, we implement several key optimizations in our web-based visualization system.

\paragraph{Parallel Task Distribution}
We leverage web workers to parallelize computationally intensive tasks while keeping the main thread responsive for user interactions:
\begin{itemize}
  \setlength{\itemsep}{0pt}
  \setlength{\parskip}{0pt}
    \item Neighbor search operations are delegated to dedicated workers that partition the search space using KD-trees
    \item Community detection algorithms run in separate workers to prevent UI blocking
    \item AMCS rendering calculations are distributed across multiple workers
\end{itemize}

\paragraph{Memory-Efficient Data Management}
To minimize memory overhead:
\begin{itemize}
  \setlength{\itemsep}{0pt}
  \setlength{\parskip}{0pt}
    \item \texttt{SharedArrayBuffer} objects enable zero-copy data sharing
    \item Binary data structures for graph representation
    \item Typed arrays for efficient numerical data storage
    \item Compression schemes leveraging nearest neighbor graph structure
\end{itemize}

These optimizations enable our system to handle datasets with hundreds of thousands of segments while maintaining interactive frame rates ($>$30 FPS) during visualization and exploration tasks.

\paragraph{Pipeline Synchronization}
The Data Processing and Computation (Figure~\ref{fig:pipeline}a) and User Interface and Interaction (Figure~\ref{fig:pipeline}b) of our pipeline are tightly synchronized. When users update parameters (control panel settings, visualization options, community detection parameters) or trigger actions like re-running community detection, these interactions initiate processes in part (a). This part recomputes necessary data (CSNG construction, community detection, rendering attributes), then passes updated data back to part (b), where visualizations (3D view, force-directed graph, adjacency matrix) update accordingly. This closed loop ensures user interactions in part (b) directly influence the computational backend in part (a), which provides updated data for further visualization and interaction in part (b).

A comprehensive demonstration of our framework and use cases is in the supplemental video.

%% file: Content/results.tex
\section{Results and Evaluation}
\label{sec:results}

We evaluate our approach using benchmark datasets including streamlines from flow behind a square cylinder, Plume simulation, Crayfish simulation, and vortex lines from turbulent Couette flow.

\subsection{Performance Analysis}
\label{subsec:performance}

Table~\ref{tab:combined_stats_performance} presents the dataset statistics and computational performance metrics across our test datasets. The performance metrics shown are execution times in milliseconds (ms) for the CSNG construction and community detection (using Louvain) and clustering (using PCA k-means). All measurements were conducted on a system equipped with an AMD Ryzen 5 3600 processor and 48GB DDR4-3600 RAM.

\begin{table}[htbp] 
\centering
\caption{Dataset statistics and performance metrics. Dataset statistics show the number of streamlines and segments. Performance metrics are execution times in milliseconds (ms) for Louvain and PCA analysis across different levels (strlines, subcurv, seg - representing streamlines, sub-curves, and segments, respectively). CSNG construction parameters: kNN ($k{=}60$), RBN (R=10\% of dataset diagonal). Louvain resolution=1.0 for all tests. 
The 222,078 ms ($\approx$222 seconds or $\approx$3.7 minutes) for CF Turb streamline-level reflects the time for CSNG construction; community detection itself takes under 300 ms. Segment-level times for CF Turb and PCA on CF Turb are not available (N/A) due to computational limitations.}
\label{tab:combined_stats_performance}
\tiny 
\begin{tabular}{|l|r|r|r|r|r|r|r|r|}
\hline
\multirow{2}{*}{\textbf{Dataset}} & \multirow{2}{*}{\textbf{strlines}} & \multirow{2}{*}{\textbf{seg}} & \multicolumn{3}{c|}{\textbf{Louvain (ms)}} & \multicolumn{3}{c|}{\textbf{PCA (ms)}} \\
\cline{4-9}
& & & \textbf{strlines} & \textbf{subcurv} & \textbf{seg} & \textbf{strlines} & \textbf{subcurv} & \textbf{seg} \\
\hline
B\'enard & 864 & 40,086 & 4,235 & 1,024 & 4,442 & 3,227 & 1,284 & 3,508 \\
Crayfish & 2,048 & 125,493 & 10,400 & 5,763 & 11,216 & 16,164 & 11,420 & 31,338 \\
Cylinder & 1,372 & 85,403 & 8,390 & 4,173 & 8,784 & 8,379 & 2,215 & 19,979 \\
Plume & 864 & 221,076 & 20,683 & 12,701 & 21,651 & 22,305 & 8,709 & 61,685 \\
CF Turb & 107,431 & 1,408,283 & 222,078 & 55,965 & N/A & N/A & N/A & N/A \\
\hline
\end{tabular}
\end{table}

The Louvain algorithm is exceptionally fast compared to traditional clustering. For our largest dataset (stress-driven turbulent Couette flow with 1.4M segments), Louvain completes in under 300ms once the CSNG is constructed. Total time including CSNG construction at the streamline level is approximately 222 seconds (Table~\ref{tab:combined_stats_performance}), but construction only needs to be performed once and can be reused for multiple runs with different parameters. This represents significant improvement over PCA $k$-means clustering, requiring several minutes. Our method maintains a 4-10x performance advantage over traditional clustering approaches.

The RBN neighbor search strategy becomes computationally prohibitive for dense datasets like the turbulent Couette flow, where the unbounded neighbor count leads to exponential growth in graph size. For such cases, kNN provides better scalability while maintaining result quality.
\paragraph{Scalability Analysis}
To evaluate the scalability of our system, we conducted a systematic study using progressively larger subsets of the Solar Plume dataset, which features densely packed streamlines particularly near its core, a challenging scenario for our neighborhood-based approach. Table~\ref{tab:scalability_analysis} presents the runtime breakdown as the dataset size increases linearly from 625 to 2,306 streamlines (805K to 2.8M segments).

\begin{table}[h!]
\centering
\caption{Scalability analysis showing runtime breakdown (in milliseconds) as the number of streamlines increases. All tests use kNN with $k{=}5$ and Louvain community detection at streamline-level with resolution=1.0. OOM indicates out-of-memory error.}
\label{tab:scalability_analysis}
\tiny
\begin{tabular}{|r|r|r|r|r|r|r|}
\hline
\textbf{Streamlines} & \textbf{Seg} & \textbf{Data} & \textbf{KD-tree} & \textbf{kNN} & \textbf{Louvain} & \textbf{Comm} \\
 & & \textbf{(ms)} & \textbf{(ms)} & \textbf{(ms)} & \textbf{(ms)} & \textbf{\#} \\
\hline
625 & 805,244 & 503 & 2,162 & 31,078 & 128 & 71 \\
1,166 & 1,475,018 & 837 & 3,917 & 77,030 & 199 & 297 \\
1,671 & 2,062,060 & 1,369 & 7,507 & 119,219 & 253 & 364 \\
2,306 & 2,822,308 & 1,758 & 12,142 & 168,473 & OOM & OOM \\
\hline
\end{tabular}
\end{table}

Our study shows data processing and KD-tree construction scale approximately linearly with dataset size, while kNN neighbor search exhibits super-linear scaling due to increased search space in denser datasets. Community detection remains efficient (under 300ms) for datasets up to 2M segments. At approximately 2.8M segments, browser-imposed memory limitations cause out-of-memory (OOM) errors. This limitation stems from browser-side constraints rather than system RAM. Our test system had 48GB RAM, but browsers impose strict per-tab memory limits for security and stability. We attempted to measure memory usage using browser developer tools, but the profiling process consumed additional resources, causing tests to OOM. At this scale, the 3D interactive view becomes unresponsive, with operations like selecting communities causing prolonged freezes. These limitations are inherent to web browsers, which typically restrict memory to 2-4GB per tab regardless of system RAM and lack GPU compute shader access.

\subsection{Comparative Analysis}
\label{subsec:comparative}

We compared our Louvain-based community detection with traditional PCA $k$-means clustering, a widely used method for flow feature identification \cite{shi2021integral}. For PCA $k$-means, the input consists of curve coordinates (3D point positions). At the streamline level, curves are resampled to the same number of points and concatenated into fixed-length feature vectors. At the segment level, each segment is represented by its two endpoint coordinates. All coordinates are normalized to [0,1] based on the bounding box of a dataset before PCA. We use Euclidean distance in the PCA-reduced feature space for k-means. Principal components retain 95\% of variance (typically 10-20 components). This analysis considers performance and quality of both methods at two granularities: \textit{streamline-level}, where each curve is treated as a single entity, and \textit{segment-level}, which operates on individual segments. Our evaluation focused on feature coherence, multi-scale performance, and parameter sensitivity.

\begin{figure*}[!t]
    \centering
    \includegraphics[width=0.98\linewidth]{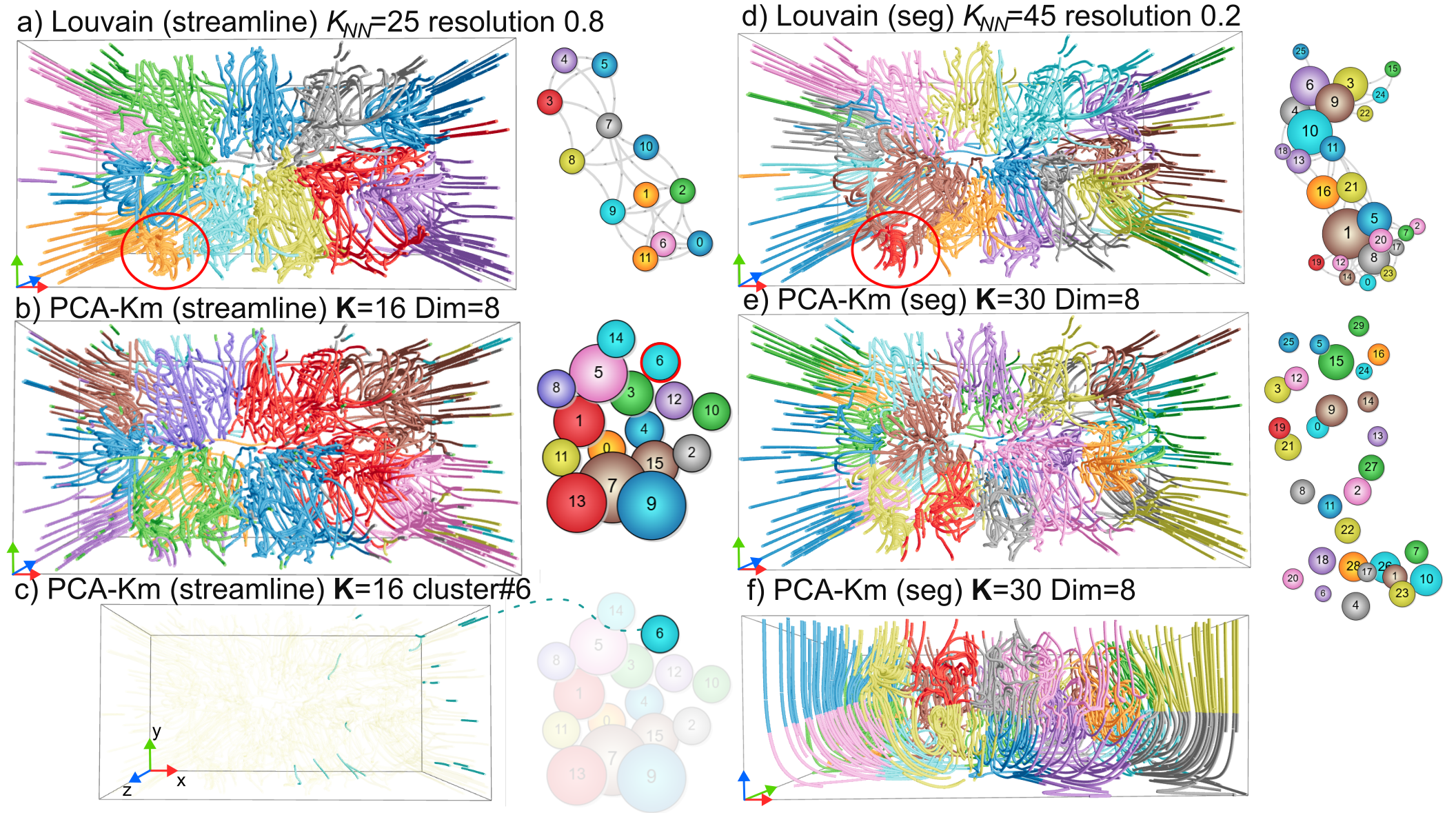}
    \caption{Comparison of Louvain and PCA $k$-means (PCA-KM for short) on the Crayfish dataset. Each column compares methods (Louvain vs. PCA-KM), while each row compares analysis levels (streamline vs. segment). 
    \textbf{First row (Louvain):} (a) Streamline-level analysis - the red circle shows how long streamlines that form the vortex feature extend to the boundary, preventing proper isolation of the local feature. (d) Segment-level analysis - the red circle shows how segment-level Louvain successfully creates a smaller community that encapsulates just the interesting vortex feature.
    \textbf{Second row (PCA-KM):} (b) Streamline-level analysis (e) Segment-level analysis
    \textbf{Third row (PCA-KM Limitations):} (c) Highlights the artifact in streamline-level PCA-KM, where short streamlines that are located away from each other are incorrectly grouped. (f) A side view of the segment-level PCA-KM result shows how clusters are discretized and do not follow natural flow structures.
    }
    \label{fig:crayfish_comparison}
\end{figure*}

\noindent\textbf{Feature Coherence.} Elements grouped into a cluster should be spatially proximate and part of the same underlying flow structure. We evaluate this on the Crayfish dataset (2048 streamlines, 227k segments), shown in ~\autoref{fig:crayfish_comparison}.

The first row shows Louvain community detection results. At the streamline level (a), it preserves spatial coherence, forming communities that align well with visual flow patterns. The red circle in (a) highlights a critical limitation: when streamlines forming a localized feature extend far beyond the feature's spatial boundaries, streamline-level analysis fails to isolate the feature properly. Here, streamlines forming the compact vortex originate from distant boundary regions, causing entire streamlines to be grouped together despite the vortex being localized. At the segment-level (d), this limitation is addressed. It successfully creates a focused community encapsulating just the vortex feature by breaking down long streamlines and allowing different portions to be assigned to appropriate local communities (see the red circle in (d)). Note that segment-level analysis generates more communities at the same resolution value, so when switching from streamline-level to segment-level, the resolution parameter should be reduced by at least 4-10 times (in ~\autoref{fig:crayfish_comparison}(a) vs (d), reduced from 0.8 to 0.2).

The second row displays PCA $k$-means results. At the streamline-level (b), it produces artifacts in laminar regions and creates discontinuous, stripe-like partitions. At the segment-level (e), it resolves the padding artifact (discussed next), but clustering quality remains suboptimal, forming groups based on proximity to centroids rather than inherent streamline features.

The third row highlights PCA $k$-means limitations. A significant issue at the streamline level is the incorrect grouping of short streamlines that are far away from each other into a single cluster (c). 
This occurs because the method requires input streamlines of identical length, forcing padding for shorter streamlines, which distorts the feature space. At the segment level, a rotated side view (f) reveals that clusters are separated based on spatial discretization rather than following natural streamline continuity. Overall, Louvain demonstrates better visual performance in capturing meaningful flow structures at both levels.

Next, we look at the turbulent Couette flow dataset~\cite{li2019direct} with 107,431 vortex lines and 1.4 million segments to evaluate our approach on a large-scale dataset with ground truth labels. The ground-truth labels were established by extracting vortical regions from the direct numerical simulation using a $\lambda_2$-based region-growing strategy~\cite{zafar2023extract}, which were then topologically separated into individual vortices~\cite{zafar2024topological}. While $\lambda_2$ identifies vortical regions, it does not automatically provide the grouping structure needed for analysis—our method addresses the complementary problem of efficiently organizing and exploring the resulting vortex line bundles for interactive visual analysis. The vortex lines were integrated from the vorticity field and labeled with the IDs of the corresponding vortices, providing ground truth for evaluation.

To quantify the accuracy of our community detection against ground truth classifications, we employ the Jaccard index, which measures the normalized overlap between predicted and actual groupings:
\begin{equation}
    J(A, B)=\frac{\sum\left(A \cap B\right)}{\sum \left( A \cup B\right)}
\end{equation}
This metric ranges from $J\in[0,1]$, where $J=0$ signifies complete disagreement between predictions and ground truth, while $J=1$ indicates perfect alignment. Our streamline-level Louvain-based approach achieved a weighted Jaccard index of 0.712, significantly outperforming PCA $k$-means at the streamline level, which achieved only 0.275, demonstrating superior alignment with the ground truth vortex bundle classifications.

\noindent\textbf{Multi-scale Performance:} 
Our framework constructs CSNGs at streamline, sub-curve, and segment levels, enabling robust multi-scale analysis. The Crayfish dataset exemplifies trade-offs between computational cost, memory usage, and ability to capture fine-grained details versus global structures. Streamline-level analysis offers significantly faster runtimes (Table~\ref{tab:combined_stats_performance}). For Louvain, aggregating neighbor connections to streamline level results in a much smaller graph. For PCA, streamline data are batched into a single square matrix, speeding up performance; however, a major limitation is the requirement that all input streamlines have the same length, necessitating padding or trimming, which introduces artifacts (~\autoref{fig:crayfish_comparison}(c)).


While segment-level Louvain allows confinement of local features, an important consideration is the dramatic increase in resulting communities. For identical kNN $k$ and resolution parameters, segment-level Louvain typically produces at least 10x more communities than streamline-level analysis, with each community significantly smaller. While this granular decomposition is valuable for hierarchical exploration of existing large communities, it presents challenges for initial dataset exploration due to the overwhelming number of small communities.

The resolution parameter in Louvain enables users to tune community granularity. However, users must be aware of the quality trade-off: decreasing resolution reduces community count but at the cost of detection quality. This trade-off is particularly pronounced in segment-level analysis, where a high community count may force aggressive resolution reduction (e.g., from 1.0 to 0.01), compromising the meaningfulness of the result. An alternative strategy to manage high community count while preserving quality is to increase the kNN $k$ parameter, which maintains community quality at lower resolution values by providing richer neighborhood information. However, this comes with high computational costs: increasing $k$ in segment-level analysis leads to substantially larger graph structures with correspondingly higher memory usage and runtime, as the segment-level graph is already much denser than the streamline level.

In contrast, PCA $k$-means shows less adaptability to inherent scale differences, as its efficacy is highly dependent on the pre-selected number of clusters ($k_c$), and fixed-length vector representation can obscure scale-dependent features. Artifacts can be observed in both levels (\autoref{fig:crayfish_comparison}(c)(f)).

\vspace{0.1in}
\noindent\textbf{Parameter Sensitivity:} Our method achieved optimal results with minimal parameter tuning. The $k$ parameter for kNN neighbor search is dataset-dependent, scaling with curve density: sparser datasets (like Plume or B\'enard) work well with $k=6{\sim}10$, while denser datasets (like Couette flow) benefit from $k=40{\sim}60$. A practical heuristic is to start with $k$ set to approximately $0.01\%$ to $0.05\%$ of the total number of elements (segments or streamlines), noting that this percentage serves only as a guideline for selecting an integer neighbor count. The resolution parameter for Louvain community detection is more stable: resolution=1.0 works well for streamline-level analysis across all datasets, while segment-level typically requires reduction to 0.05 -- 0.2 to avoid excessive fragmentation. In contrast, PCA $k$-means required careful adjustment of both $k_c$ (number of clusters) and dimensional reduction parameters (number of principal components to retain, typically 10$\sim$20) as described in previous flow visualization work \cite{shi2021integral}.

\subsection{Use Cases of Our Framework}

We apply our framework to explore two complex flow datasets: Solar Plume and stress-driven turbulent Couette flow. These use cases showcase how our approach enables automated feature detection and interactive refinement of flow structures.

\begin{figure}[!t]
  \centering
  \includegraphics[width=0.85\linewidth]{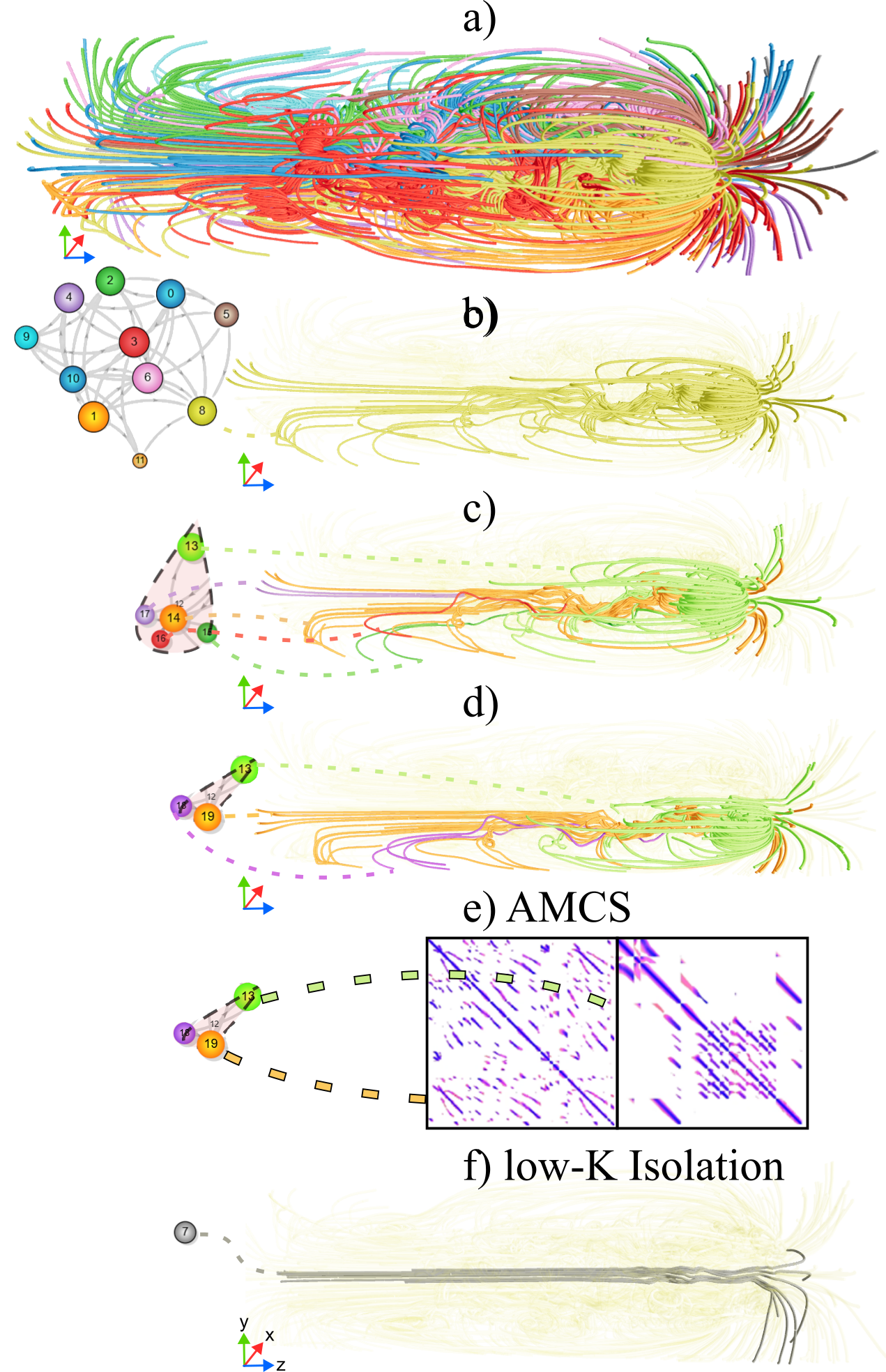}
  \caption{Analysis of the Plume dataset using hierarchical community exploration. (a) Initial 3D visualization with segments colored by Louvain communities. (b) Community \#8 highlighted, showing complex vortex features requiring further analysis. (c) Splitting of community \#8 into five subcommunities, with the largest vortex structure maintained as a single coherent subcommunity. (d) Final refined structure after manual community merging: \#14 merged with \#17, and \#15 with \#17, resulting in three well-defined subcommunities that capture distinct flow features. (e) AMCS View of communities \#19 (left) and \#13 (right), respectively, showing distinct matrix patterns. (f) Isolation of community \#7 (highlighted) reveals core streamlines, demonstrating the effectiveness of low-K neighbor search in identifying coherent flow structures.}
  \label{fig:demo}
\end{figure}

\subsubsection{Solar Plume Analysis}

We explore the Solar Plume dataset following the recommended workflow: (1) Start with streamline-level CSNG construction using kNN ($k$ set to $\sim$0.01-0.05\% of streamline count) for fast initial exploration, (2) Apply Louvain with resolution=1.0 to identify large-scale features, (3) Examine force-directed graph and 3D view to identify communities of interest, (4) Selectively switch to segment-level analysis for specific communities requiring finer detail, and (5) Use AMCS matrix view to examine detailed geometric patterns. This hierarchical workflow balances computational efficiency with analysis depth. As established in our comparative analysis, Louvain better maintains feature coherence. We constructed a CSNG using kNN ($k{=}10$) and applied streamline-level Louvain detection (resolution=1.0).

Initial community detection revealed several distinct flow features (Figure~\ref{fig:demo}(a)). Our analysis focused on community\#8 (Figure~\ref{fig:demo}(b)), which contained multiple complex vortex features. To refine our understanding, we switched to segment-level analysis on just this community. This hierarchical approach (moving from coarse streamline-level overview to fine-grained segment-level examination) allows efficient and detailed exploration without the high upfront cost of global segment-level analysis. This splitting divided the region into five subcommunities (Figure~\ref{fig:demo}(c)). We performed manual community merging, resulting in three well-defined subcommunities (Figure~\ref{fig:demo}(d)) that better represent the natural hierarchical structure. 
Next, we inspect two of these subcommunities, i.e., \#19 and \#13, using their respective AMCS views (Figure~\ref{fig:demo}(e)). They exhibit distinct matrix patterns: \#19 (left matrix) exhibits long parallel diagonal lines, indicating that the segments in \#19 are mostly pointing in the same direction. In contrast, \#13 (right matrix) shows dense multi-layered diagonal patterns, indicating swirling (or vortex-like) features. This provides granular configurations of segments within a community that the force-directed layout cannot effectively convey.

A particularly interesting finding emerged in community\#7 (Figure~\ref{fig:demo}(f)) of the above example, which appeared completely isolated in the force-directed graph. The line set in Figure~\ref{fig:demo}(f) is a subset of input streamlines (only those belonging to community \#7) displayed to highlight this isolated feature. This isolation captured the core plume streamlines, which exhibit highly linear behavior distinct from surrounding turbulent flow. To enhance the identification of distinctive flow features, our framework incorporates opacity threshold filtering based on community properties. The system provides three predefined template filter settings: (1) \textit{large isolated communities} (identifying spatially separated structures like core plume streamlines in Figure~\ref{fig:demo}(f)), (2) \textit{large communities with high average distance} (highlighting dispersed patterns spanning significant spatial regions), and (3) \textit{small communities with high connectivity} (detecting compact, highly interconnected structures indicating interesting features such as vortex cores or recirculation zones). These small, highly connected communities are noteworthy because their persistence as separate entities despite high connectivity suggests distinct physical phenomena. Users can directly select from these predefined templates or manually configure custom filter pipelines through rendering settings.

\subsubsection{Analysis of Stress-Driven Turbulent Couette Flow Dataset}

The stress-driven turbulent Couette flow dataset (CF) \cite{li2019direct} presents a particularly challenging test case. The vortex lines -- streamlines of the vorticity vector field -- form dense bundles that are sparsely distributed, creating regions with varying vortex density across the domain. Here, a global streamline-level analysis is still the most efficient starting point.

\begin{figure*}[!t]
    \centering
    \includegraphics[width=0.95\linewidth]{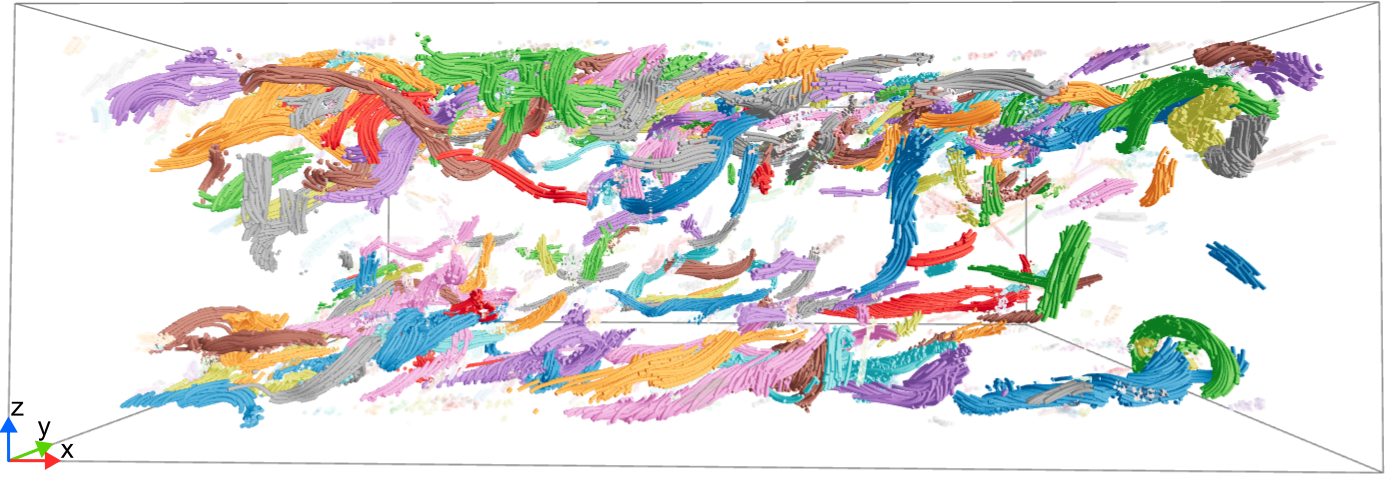}
    \caption{The actual Louvain communities result demonstrates how the method effectively encapsulates vortex bundles into coherent communities, with each color representing a distinct community that closely follows the natural flow structures.}
    \label{fig:louvain_large}
\end{figure*}

\begin{figure}[!t]
    \centering
    \includegraphics[width=0.99\linewidth]{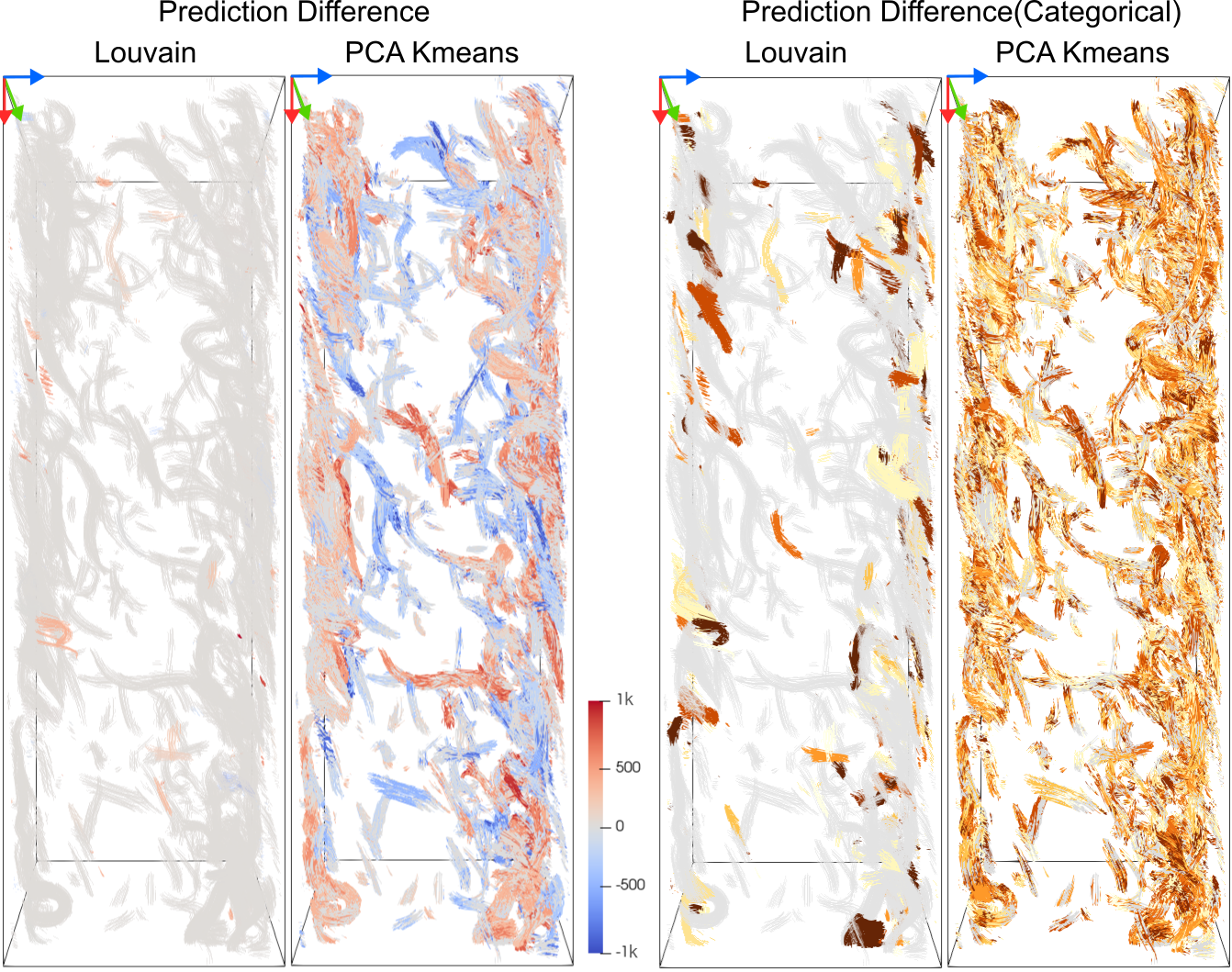}
    \caption{Quantitative comparison between Louvain and PCA $k$-means on the CF dataset. Both sub-figures show the prediction difference, calculated as the difference between the predicted and ground-truth vortex IDs.
    \textbf{Left (Continuous Difference):} This view uses a continuous color scale where blue indicates a higher predicted index and red indicates a lower one, relative to the ground truth. Gray signifies a very close match. While this shows Louvain performs well, the proximity of some mismatched indices can make them appear gray, suggesting a better match than is strictly the case.
    \textbf{Right (Categorical Difference):} This view highlights any mismatch, regardless of how close the indices are. Each distinct color represents a unique vortex, making it clear where classifications are incorrect. This provides a stricter evaluation.
    Louvain's predominantly gray visualization in the left panel and more consistent coloring in the right indicate better alignment with ground truth (weighted Jaccard index: 0.712) compared to PCA $k$-means (weighted Jaccard index: 0.275), which exhibits higher errors.}
    \label{fig:jh_comparison_quant}
\end{figure}

We evaluated both segment-based and streamline-based Louvain approaches on this dataset. The streamline-based approach ($k{=}6$, resolution=1.0) completed in just 294.7ms and yielded more coherent communities with a Jaccard index 0.712 (Figure~\ref{fig:jh_comparison_quant}), with visual inspection confirming communities closely matched ground truth vortex bundles. Ground-truth labels were established by extracting vortical regions from direct numerical simulation~\cite{li2019direct} using $\lambda_2$-based region-growing~\cite{zafar2023extract}, then topologically separated into individual vortices~\cite{zafar2024topological}. Vortex lines were integrated from the vorticity field and labeled with corresponding vortex IDs. Louvain results demonstrate good visual quality, effectively encapsulating vortex bundles into coherent communities (Figure~\ref{fig:louvain_large}). Our framework can render this dataset with 1.4M segments in real time for interactive exploration, enabling users to dynamically investigate community structures. In contrast, PCA $k$-means takes over 20 minutes for 1000 clusters and 100 iterations, a major challenge for real-time interactivity. This highlights the power of our recommended workflow and the robustness of Louvain over PCA-based methods, which struggle with density variations.

A significant limitation with this dense dataset is the large number of resulting communities (approx. 1,300 for streamline-based detection). This high count poses visualization challenges for the force-directed graph, which becomes unusable beyond 200 communities. While specific to the Couette flow dataset, our method remains effective for datasets like Plume or Crayfish where analysis can begin with fewer, larger communities. As shown in Figure~\ref{fig:demo}(b), the force-directed layout can effectively highlight user-selected communities, providing an intuitive interface for exploring dataset structure.

\subsubsection{Improvements Over Previous Work}
\label{sec:improvements_moved}
This work substantially extends previous work~\cite{phan2024curve} in several key aspects. First, we significantly enhanced computational efficiency and memory footprint. Our Louvain-based implementation now performs several orders of magnitude faster than comparable methods such as PCA-$k$-means, coupled with more efficient memory usage, particularly at the segment level. For example, on the Cylinder dataset with over 30,000 segments, our method completes in under 20 seconds for segment-level Louvain, while PCA-$k$-means takes substantially longer and consumes more memory. 
Table~\ref{tab:runtime_comparison} provides detailed performance gains across datasets compared with previous work~\cite{phan2024curve}. The comparison is conducted at the segment level with $k{=}25$ for kNN. Additionally, previous work~\cite{phan2024curve} did not support streamline-level analysis, now a key capability of our enhanced framework. These improvements are crucial for enabling interactive exploration of large-scale data.

\begin{table}[h!]
\centering
\caption{Comparison of Old~\cite{phan2024curve} and New Louvain Runtimes and Memory Usage ($k{=}25$ for segment-level CSNG)}
\label{tab:runtime_comparison}
\small
\begin{tabular}{|l|cc|cc|}
\hline
\textbf{Dataset} & \multicolumn{2}{c|}{\textbf{Runtime (s)}} & \multicolumn{2}{c|}{\textbf{Memory (MB)}} \\
\cline{2-5}
 & \textbf{Old} & \textbf{New} & \textbf{Old} & \textbf{New} \\
\hline
B\'enard & 6.38 & \textbf{4.44} & 946.9 & \textbf{488.8} \\
Crayfish & 20.54 & \textbf{11.22} & 1824.5 & \textbf{1171.4} \\
Cylinder & 11.80 & \textbf{8.78} & 1425.3 & \textbf{821.6} \\
Plume & 45.03 & \textbf{21.65} & 2738.5 & \textbf{1491.5} \\
\hline
\end{tabular}
\end{table}

Second, we introduce the Adjacency Matrix of Curve Segments (AMCS) visualization, supporting detailed manual refinement of community results (step 4 of our analysis pipeline, Figure~\ref{fig:pipeline}). AMCS provides a powerful complementary view to our force-directed layout. This matrix representation enables users to identify subtle patterns and relationships between curve segments that might be obscured in force-directed visualization alone, particularly in regions of high curve density (e.g., Figure~\ref{fig:demo}(e) shows AMCS revealing patterns in Plume dataset exploration).


These improvements collectively enable more robust and comprehensive analysis of complex flow datasets, particularly in scenarios involving high curve density where traditional methods struggle to provide meaningful insights.

%% file: Content/conclusion.tex
\section{Discussion and Conclusion}
\label{sec:discussion}

We have presented a comprehensive framework for visualizing and exploring 3D streamlines through the Curve Segment Neighborhood Graph (CSNG). Our approach combines efficient community detection algorithms with multi-level visualization techniques, including force-directed layouts and the Adjacency Matrix of Curve Segments (AMCS). Our framework offers a fast, scalable, and qualitatively superior solution for streamline analysis compared to traditional methods like PCA $k$-means. The key to its effectiveness is the flexibility to analyze data at both the streamline and segment levels, enabling a hierarchical exploration workflow. Our findings suggest a general strategy: start with the faster, more memory-efficient streamline-level analysis to get an overview, then use the more detailed segment-level analysis on specific regions of interest. This balances performance with analytical depth. Nevertheless, the framework is flexible enough to start with segment-level analysis when complex streamline configurations demand it.

A particularly significant discovery is the effectiveness of low-$k$ kNN configurations ($k \leq 10$) for early-stage feature identification, as they naturally yield sparser graphs that better isolate coherent structures. In the Solar Plume dataset, using $k=10$ led to community \#7 appearing fully separated in the force-directed layout (~\autoref{fig:demo}(f)), capturing the plume’s highly linear core distinct from the surrounding turbulence. This separation highlights how low-$k$ graphs can emphasize physically meaningful regions without prior knowledge of the flow.


\subsection{Limitations}

While our web-based framework provides exceptional accessibility—-running on any device with a browser and offering a responsive React interface—-this design choice involves significant performance trade-offs. The system suffers from limited multithreading capabilities, where web workers add overhead to runtime, hard memory limits that prevent utilizing full system memory, and the need to avoid UI freezing by delegating expensive tasks to web workers, which introduces data transfer overhead and garbage collection variability. Large operations like community detection or neighbor search can vary by 10-20\% in runtime with no external factors, and these limitations become more pronounced as the size of the data increases.

Our system focuses on analyzing individual time steps or steady-state flow fields. We do not currently support temporal tracking of communities across time steps or ensemble analysis. While our real-time performance could enable frame-by-frame analysis, extending CSNG to encode temporal adjacency relationships would enable automatic analysis across temporal dimensions. Additionally, the interactive split/merge operations introduce subjectivity, as different users may refine communities differently based on their domain knowledge. This is intentional for exploratory analysis and hypothesis generation, but users requiring reproducible results should rely on the automatic community detection baseline. Finally, unlike topology-based feature extraction methods, our neighborhood-based approach does not guarantee identification of all structures of interest—-detected communities depend on neighbor search parameters and resolution settings. We recommend exploring multiple parameter configurations and validating results with domain knowledge or quantitative metrics when available.

\subsection{Future Work}

While our current implementation shows considerable promise, we have identified several directions for future research. The impact of different neighbor search strategies (RBN versus kNN) on CSNG construction and subsequent analysis requires further investigation. We plan to explore more sophisticated curve-centered neighbor search methods to address the limitations of current approaches and reduce parameter sensitivity. Additionally, we aim to enhance the handling of local patterns and global structures by investigating alternative community detection algorithms and exploring methods for encoding multi-scale flow information into CSNGs.

Our future work will also focus on improving the scalability of our web-based implementation, particularly for processing datasets with a very large number of segments. To achieve that, future work will explore a dedicated server-client architecture where computationally intensive operations (CSNG construction, community detection) are offloaded to a high-performance server, while the client browser handles interactive visualization using techniques such as server-side rendering (SSR) and progressive data streaming. This approach would enable our framework to scale beyond current browser limitations while maintaining the accessibility and cross-platform benefits of web-based deployment. We plan to explore WebGPU acceleration for computationally intensive operations, though Louvain and $k$-means remain fundamentally CPU-bound methods. Finally, we will conduct more thorough comparisons between our community detection approach and conventional clustering methods to better understand their relative strengths and limitations across different types of flow datasets.